\documentclass{nic-series}
\usepackage[pdftex]{color}

\usepackage{epsf}

\newcommand{\aap}{Astron. \& Astrophys.}
\newcommand{\apj}{Astrophys. J.}
\newcommand{\apjl}{Astrophys. J., Letters}
\newcommand{\apjsup}{Astrophys. J. Suppl. Ser.}
\newcommand{\apss}{Astrophysics and Space Science}
\newcommand{\nature}{Nature}
\newcommand{\nat}{Nature}
\newcommand{\prd}{Phys. Rev. D}
\newcommand{\prl}{Phys. Rev. Lett.}
\newcommand{\prep}{Phys. Rep.}
\newcommand{\pre}{Phys. Rev. E}

\newcommand{\mnras}{Mon. Not. R. Astron. Soc.}
\newcommand{\jkas}{Journ. Korean. Astron. Soc.}
\newcommand{\araa}{Ann. Rev. Astron. Astrophys.}
\newcommand{\znat}{Z. Naturforsch}

\newcommand{\sm}[1]{\rm{{\scriptsize #1}}}

\begin{document} 

\title{Generation of strong magnetic fields via the small-scale dynamo during the formation of the first stars}

\author{Robi Banerjee \inst{1}, Sharanya Sur \inst{2,3}, Christoph Federrath \inst{3,4,5}, Dominik R. G. Schleicher \inst{6}, Ralf S. Klessen \inst{3}}

\institute{Hamburger Sternwarte, Universit\"at Hamburg, \\
         Gojenbergsweg 112, 21029 Hamburg, Germany\\
         \email{banerjee@hs.uni-hamburg.de}
         \and
         Inter-University Centre for Astronomy and Astrophysics, \\
         Post Bag 4, Ganeshkhind, Pune-411007, India
         \and
         Zentrum f\"ur Astronomie der Universit\"at Heidelberg, \\
         Institut f\"ur Theoretische Astrophysik, \\
         Albert-Ueberle-Str. 2, 69120 Heidelberg, Germany \\
         \email{ \{sur, federrath, klessen\}@uni-heidelberg.de} 
         \and
         Ecole Normale Sup\'erieure de Lyon, \\
         CRAL, 69364 Lyon Cedex 07, France
         \and
         Monash University, Victoria 3800, Australia 
         \and
         Institut f\"ur Astrophysik, Georg-August-Universit\"at G\"ottingen, \\
         Friedrich-Hund-Platz 1, 37077 G\"ottingen, Germany \\
         \email{dschleic@astro.physik.uni-goettingen.de}
}

\maketitle


\begin{abstracts}
  Here we summarize our recent results of
  high-resolution computer simulations on the turbulent amplification
  of weak magnetic seed fields showing that such fields 
  will be exponentially amplified also during the gravitational collapse
  reminiscent to the situation during primordial star formation.
  The exponential magnetic field
  amplification is driven by the turbulent small-scale dynamo that
  can be only observed in computer simulations if the turbulent
  motions in the central core are sufficently resolved. We find that
  the Jeans length, which determines the central core region, has to
  be resolved by at least 30 grid cells to capture the
  dynamo activity.
  We conclude from our studies that strong magnetic fields will be
  unavoidably created already during the formation of the first stars
  in the Universe, potentially influencing their evolution and mass
  distribution.
 \end{abstracts}

\section{Introduction}

Magnetic fields are ubiquitous in the local Universe \cite{Beck+96}
and there is growing evidence of their presence also at high redshifts
\cite{B+08, RQH08, Murphy09}. The seeds for these fields could be a
relic from the early Universe, possibly arising due to inflation or
some other phase transition process \cite{TW88}. Alternatively, they
could be generated by the Biermann battery \cite{B50, Xu+08} or the
Weibel instability \cite{SS03, M+04}. Regardless of their physical
origin, most models predict weak field strengths and/or 
have large uncertainties \cite{GR01,
  BJ04}. Consequently, magnetic fields are thought to be irrelevant
for the formation of the first stars and galaxies \cite{Bromm09}. Here
we show that strong and dynamically important magnetic fields will be
generated from weak initial magnetic seed fields in the presence of
the small-scale turbulent dynamo during the collapse of primordial
halos \cite{Sur10, Federrath11a}. The properties of the small-scale
dynamo have been explored both in computer simulations of driven
turbulence without self-gravity and in analytic models
\cite{Haugen04b, Schekochihin04, BS05, Federrath11b} as well as in the
context of magnetic fields in galaxy clusters \cite{Xu09}. Analytic
estimates show that the small-scale dynamo could be important already
during the formation of the first stars and galaxies
\cite{Arshakian09,
  Schleicher10a, SO10}.

Field amplification via the small-scale dynamo requires both turbulent
gas motions and high magnetic Reynolds numbers. The presence of such
turbulence is suggested by cosmological hydrodynamical simulations of
first star formation \cite{ABN02, BCL02, YOH08} where it also plays an
important role in regulating the transport of angular momentum. High
magnetic Reynolds numbers are expected in primordial gas as it follows
closely the conditions of ideal magnetohydrodynamics (MHD).


\section{Numerical method and initial conditions}

In this study, we focus on the gravitational collapse and magnetic
field amplification of the inner parts of a contracting primordial
halo. The initial conditions for our computer simulation were
motivated from larger-scale cosmological models \cite{ABN02,
  BCL02,YOH08}. Thus, we set up a super-critical Bonnor-Ebert (BE)
sphere with a core density of $\rho_{\sm{BE}} \simeq 4.68\times 10^{-20}\,
{\rm g\, cm}^{-3}$ ($n_{\sm{BE}} = 10^4\,{\rm cm}^{-3}$) and a small amount
of rotation, $E_{\rm rot}/E_{\rm g} = 0.04$. We use a random initial
velocity field with transonic velocity dispersion, and a weak random
magnetic field with $B_{\rm rms} \sim 1 {\rm nG}$. Both the turbulent
energy and magnetic field spectra were initialized with the same power
law dependence, $\propto k^{-2}$, with most power on large
scales. Consistent with previous studies that follow the
thermodynamics during the collapse, we adopt an effective equation of
state with $\gamma=d\log T / d\log\rho + 1 = 1.1$ for all densities
relevant to this study \cite{Omukai05, GS09}. For this setup, we solve
the equations of ideal magnetohydrodynamics (MHD) including
self-gravity with an adaptive-mesh refinement technique
\cite{Fryxell00}, which guarantees that the critical length scale of
gravitational collapse is always resolved with the same number of
cells regardless of density. This scale, the local Jeans length
\begin{equation}
\label{eq:jeanslength}
\lambda_{\rm J} = \left(\frac{\pi\,c_{\rm s}^{2}}{G\,\rho}\right)^{1/2}\,,
\end{equation}
where $c_{\rm s}$ and $G$ are sound speed and gravitational constant,
is set by the competition between gravity and thermal pressure. This
is also the turbulent injection scale below which the small-scale
dynamo is active. We note that the efficiency of the dynamo process
depends on the Reynolds number and is thus related to how well the
turbulent motions are resolved \cite{HBD04}. Higher resolution results
in larger field amplification. To demonstrate this effect, we perform
five computer simulations where we resolve $\lambda_{\rm J}$ by either
8, 16, 32, 64 or 128 cells. Our results indicate that a minimum
resolution of 30 cells per Jeans length is required to see an
exponential growth of the magnetic field. 

\section{Results}

\subsection{Physical properties}

\begin{figure}
\begin{center}
\includegraphics[width=0.47\linewidth]{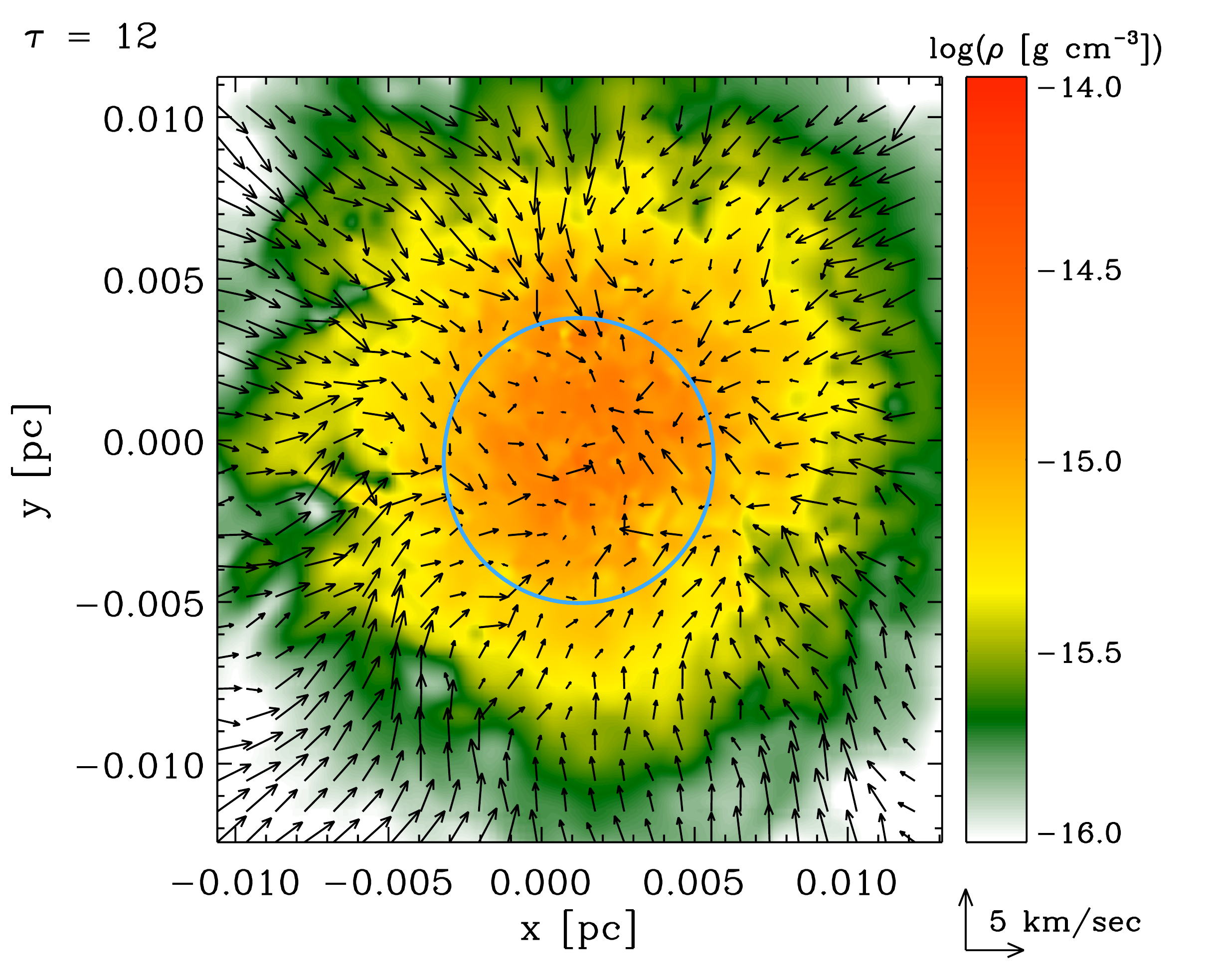}
\includegraphics[width=0.47\linewidth]{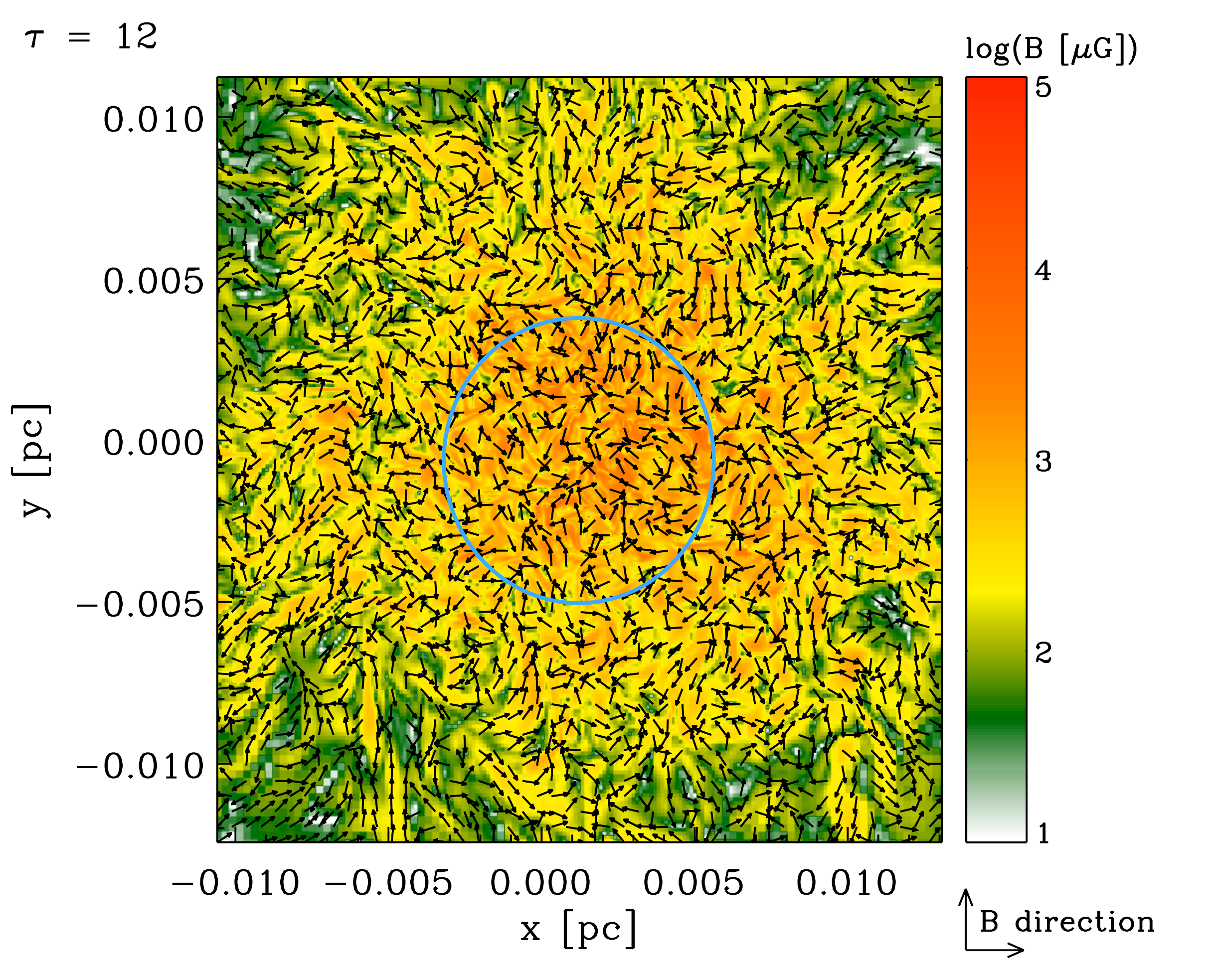}
\end{center}
\caption{Two-dimensional slices through the center of the collapsing core at the time when the initial field strength has increased by a factor of $\sim\!10^6$, showing the central region of about $0.02\times0.02\,{\rm pc}^2$ in size for our highest-resolution simulation ($\lambda_{\rm J}$ resolved by 128 cells). The circle indicates the central averaging volume $V_{\rm J}$. The left image shows the density and the velocity component in the $xy$-plane, indicating radial infall in the outer regions and turbulent motions in the inner core. The right image depicts the total magnetic field amplitude and direction. [Figure taken from \cite{Sur10}] 
  \label{fig:slices}}
\end{figure}

\begin{figure}
\centerline{\includegraphics[width=0.7\linewidth]{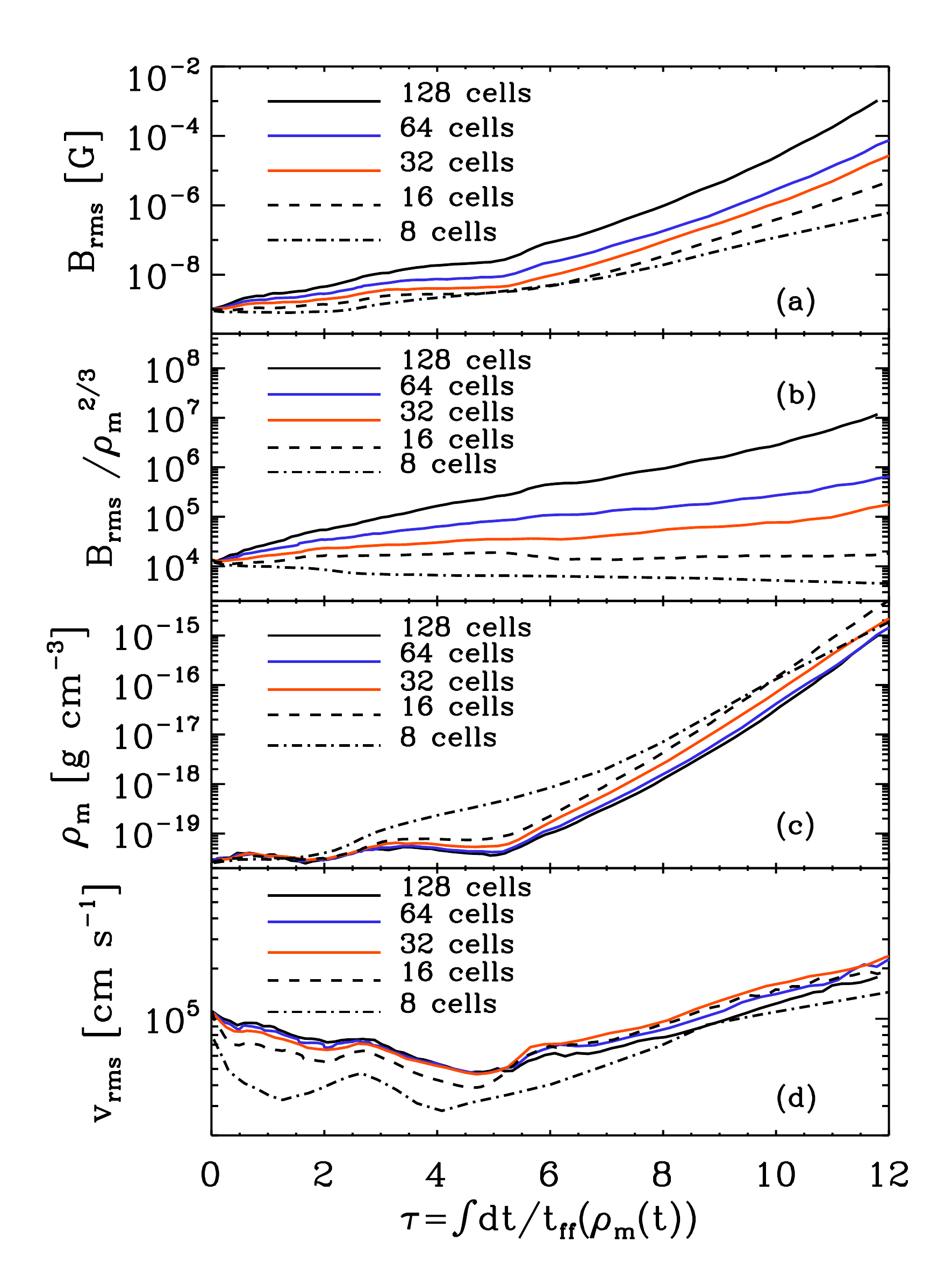}}
\caption{Evolution of the dynamical quantities as a function of $\tau = \int\,dt/t_{\rm ff}$, 
defined in equation~(\ref{tau})
for five runs with different number of cells to resolve the local Jeans length. 
Panel (a) - the rms magnetic field strength $B_{\rm rms}$, amplified to $1\,$mG from 
an initial field strength of $1\,$nG, (b) - the evolution of 
$B_{\rm rms}/\rho_{\rm m}^{2/3}$, showing the turbulent dynamo
amplification by dividing out the maximum possible amplification due to pure
compression of field lines, (c) - the evolution of the mean density 
$\rho_{\rm m}$ and (d) - the rms velocity $v_{\rm rms}$. The onset of runaway 
collapse commences at about $\tau\sim 6$. [Figure taken from \cite{Sur10}] 
\label{fig:evol}}
\end{figure}

\begin{figure}
\begin{center}
\includegraphics[width=0.49\linewidth]{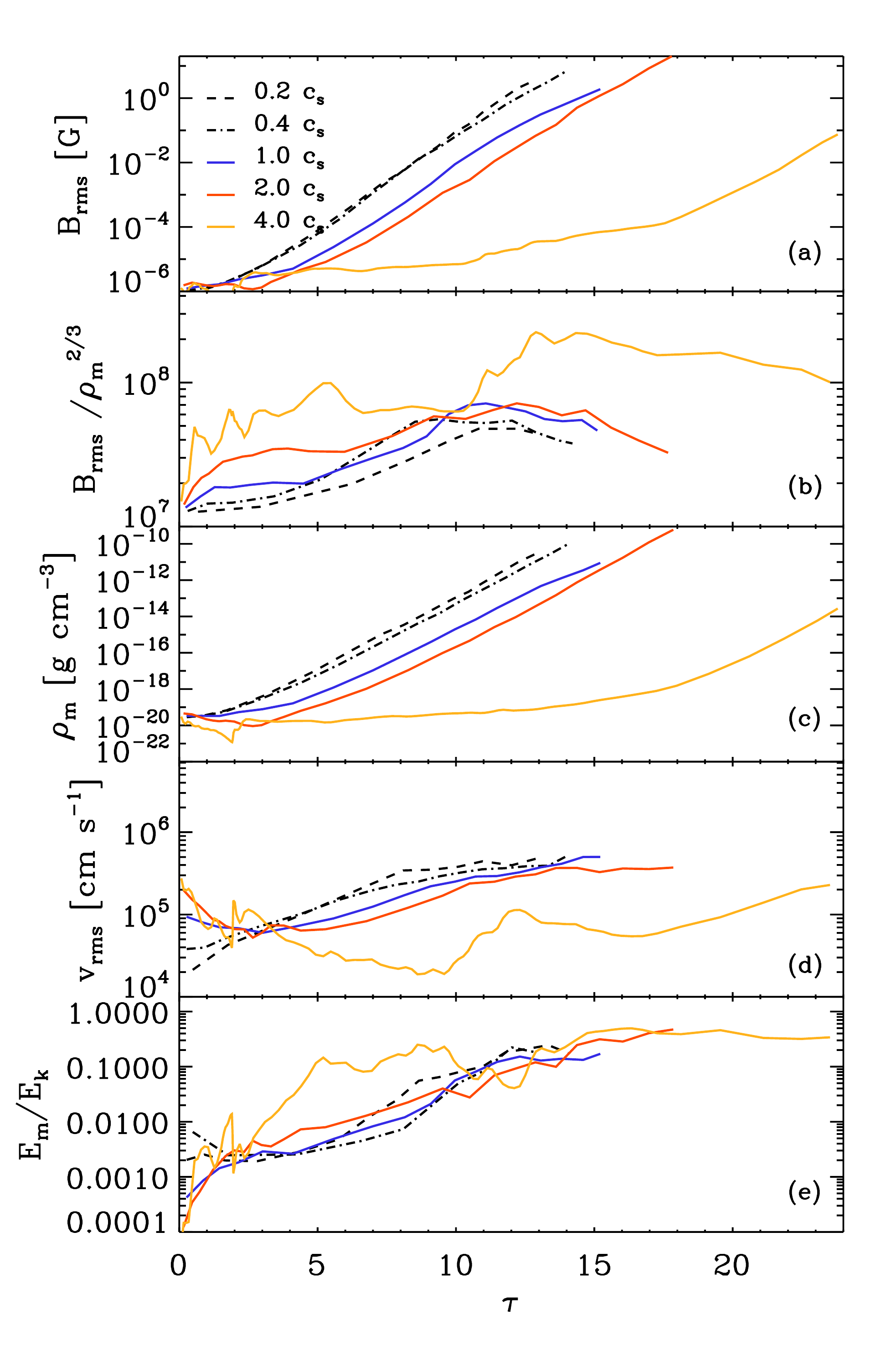}
\includegraphics[width=0.49\linewidth]{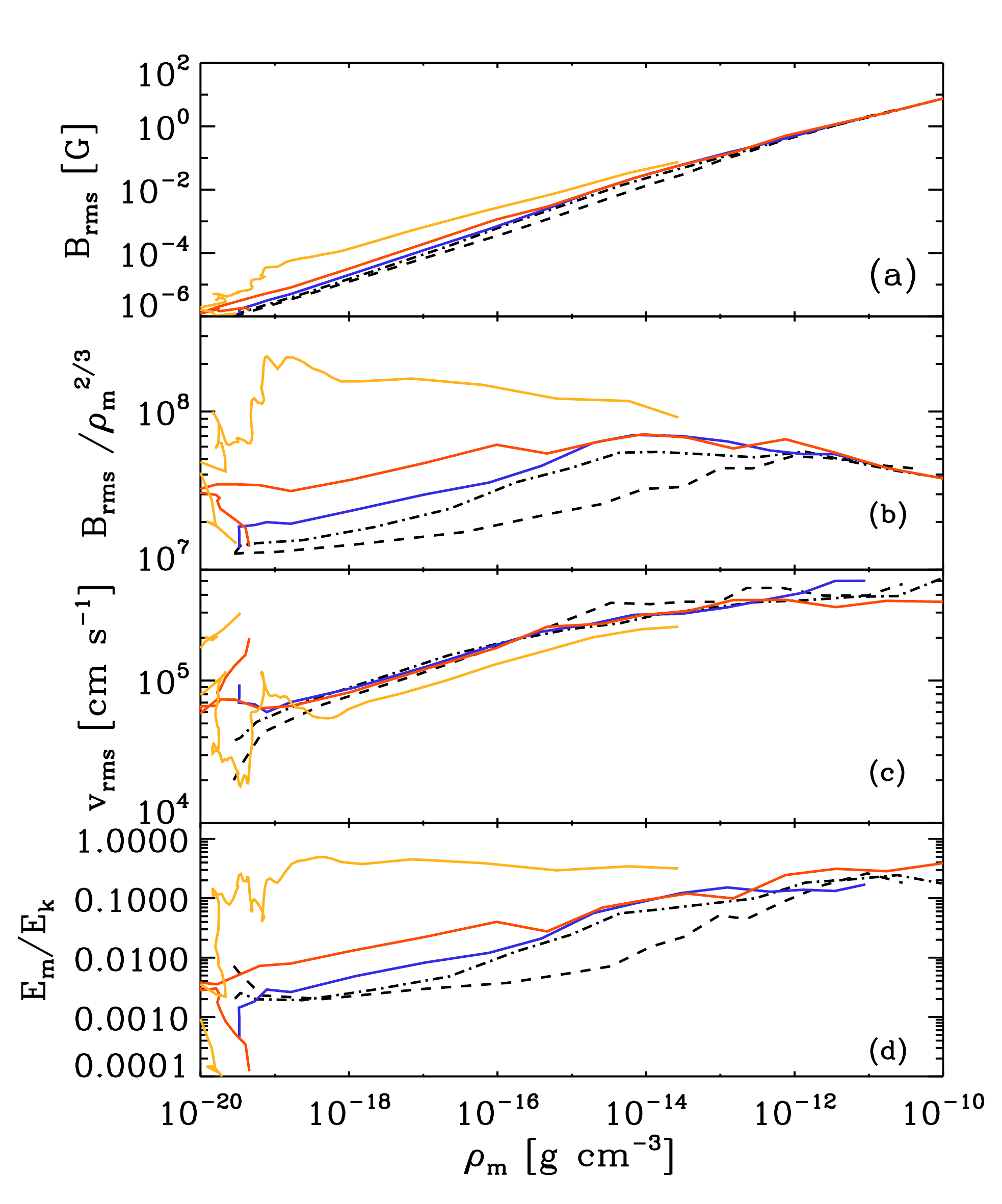}
\end{center}
\caption{Evolution of the dynamical quantities in the case of stronger initial fields ($B = 10^{-6}$ G) within the central Jeans volume as a function of $\tau$ (left) and the mean density $\rho_m$ (right) for runs with different initial Mach numbers. Shown are the rms magnetic field strength $B_{\sm{rms}}$, the ratio $B_{\sm{rms}}/\rho^{2/3}_m$, the mean density (only in the left panel), the rms velocity $v_{\sm{rms}}$ and the ratio of magnetic to kinetic energy, $E_{\sm{m}}/E_{\sm{k}}$. All the simulations correspond to a resolution of 128 cells per Jeans length.
[Figures taken from \cite{Sur11}] 
\label{fig:saturation}}
\end{figure}

We find that the dynamical evolution of the system is characterized by
two distinct phases. First, as the initial turbulent velocity field
decays the system exhibits weak oscillatory behavior and contracts
only slowly. Soon, however, the run-away collapse sets
in. Figure~\ref{fig:slices} shows a snapshot of the central region of
the collapsing core in our highest-resolution simulation at a time
when the central density has increased by a factor of $\sim 10^5$. The
magnetic field strength has grown by a factor of $10^6$, reaching peak
values of about $1\,{\rm mG}$. The left image shows density and
velocity structure, and the right image shows magnetic field strength
and morphology.

To understand the behavior of the system more quantitatively, we need
to account for its dynamical contraction. First, we note that the
physical time scale becomes progressively shorter during collapse. We
therefore define a new time coordinate $\tau$,
\begin{equation}
\label{tau}
\tau = \int dt / t_{\rm ff} (t)\,,
\end{equation}
based on the local free-fall time, $t_{\rm
  ff}(t)=\sqrt{3\pi/(32G\rho_{\rm m}(t))}$, where $\rho_{m}(t)$ is the
mean density of the contracting central region. We can also define a
critical volume for gravitational collapse, the so-called Jeans
volume, $V_{\rm J} = 4\pi (\lambda_{\rm J}/2)^3/3$, with $\lambda_{\rm
  J}$ given by equation~(\ref{eq:jeanslength}). We obtain all
dynamical quantities of interest as averages within the central Jeans
volume. This approach ensures that we always average over the relevant
volume for collapse and field amplification.

Gravitational compression during the collapse of a primordial gas
cloud can at most lead to an amplification of the magnetic field
strength by a factor of $\sim \rho^{2/3}$ in the limit of perfect flux
freezing (i.e., ideal MHD). A stronger increase implies the presence
of an additional amplification mechanism. Starting from an initial
field strength of $\sim 1{\rm nG}$, our simulations show a total
magnetic field amplification by six orders of magnitude, leading to a
field strength of about $\sim 1\,$mG for the case where we resolve the
local Jeans length by 128 cells. This is illustrated in
Fig.~\ref{fig:evol}a. Fig.~\ref{fig:evol}b shows that the obtained
field amplification is indeed stronger than what is expected for
purely adiabatic compression, which demonstrates that the small-scale
turbulent dynamo provides significant additional field amplification
over compression. A closer comparison of Fig.~\ref{fig:evol}a
and~\ref{fig:evol}b shows that the amplification of the field by
compression and by the turbulent dynamo are roughly comparable. The
time evolution of the central density $\rho_{\rm m}$ is depicted in
Fig.~\ref{fig:evol}c, while Fig.~\ref{fig:evol}d shows the
corresponding rms velocity. The presence of turbulence delays the
collapse until $\tau\sim 6$. During this time, the mean density shows
some oscillations while the rms velocity decreases as the turbulence
decays. A comparison between Fig.~\ref{fig:evol}b and
Fig.~\ref{fig:evol}d indicates that dynamo amplification takes place
throughout the entire duration of the simulation, that is during the
initial phase of turbulent decay, i.e., for $\tau \lesssim 6$, as well
as during the run-away collapse phase ($\tau \gtrsim 6$). We note that
the time constant for the field amplification is roughly the same in
both regimes.

\subsection{Implications for numerical resolution}

\begin{figure}
\begin{center}
\includegraphics[width=0.9\linewidth]{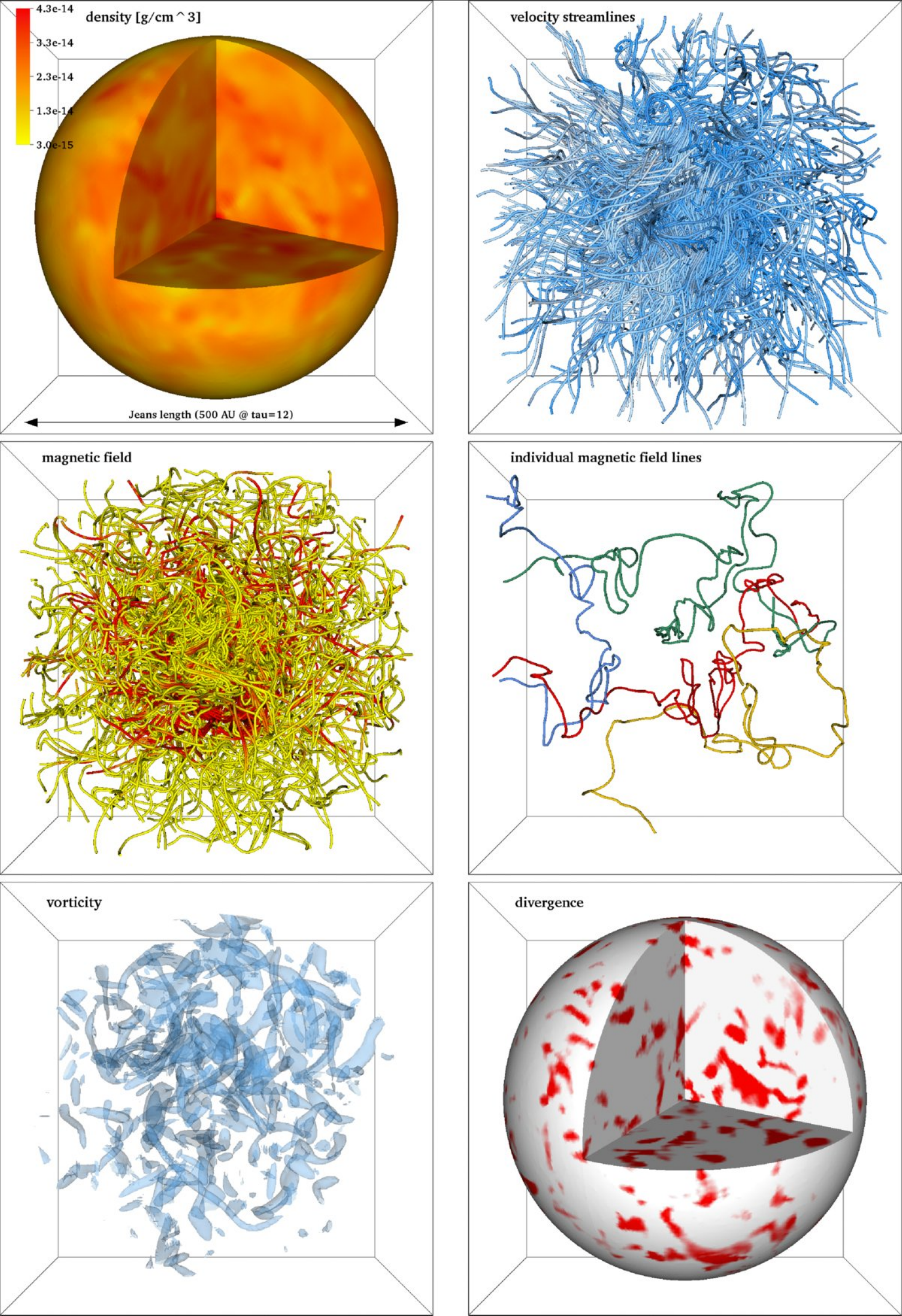}
\end{center}
\caption{ (a) Spherical slice of the gas density inside the Jeans
  volume at $\tau = 12$ for our run with 128 cells per Jeans length.
  (b) Velocity streamlines on a linear color scale ranging from dark
  blue (0 km s${}^{-1}$) to light gray (5 km s${}^{-1}$). (c) Magnetic
  field lines, showing a highly tangled and twisted magnetic field
  structure typical of the small-scale dynamo; yellow: 0.5 mG, red: 1
  mG. (d) Four randomly chosen, individual field lines. The green one,
  in particular, is extremely tangled close to the center of the Jeans
  volume. (e) Contours of the vorticity modulus, showing elongated,
  filamentary structures typically seen in subsonic turbulence. (f)
  Spherical slice of the divergence of the velocity field, white:
  compression, red: expansion. [Figure taken from \cite{Federrath11a}]}
\label{fig:structure}
\end{figure}

We also point out, that sufficient numerical resolution is a crucial
issue when studying the small-scale turbulent dynamo. We see that a
minimum of 30 cells per Jeans length is required to unambiguously
identify the exponential growth of the magnetic field
\cite{Federrath11a, Federrath11b}.
Seeing the dynamo process is computationally extremely demanding, and
even our $128^3$ high-resolution run is by no means numerically
converged. This reflects the fact that no numerical scheme to date is
able to reach the enormous Reynolds numbers, which determines the
dynamo growth rate, of star-forming, turbulent gas. Therefore, this
simulations show no signs of saturation, thus we simply stop the
calculation when the numerical cost becomes prohibitively high. In our
subsequent study \cite{Sur11} we investigate the saturation behavior
when back-reactions of the Lorentz force become important. Here we
find typical values of $E_{\sm{mag}}/E_{\sm{kin}}$ of $0.2 - 0.4$
(see Fig.~\ref{fig:saturation}) in agreement with calculations of
non-selfgravitating MHD turbulence \cite{S99, Federrath11b}. The
physical dissipation scales are much smaller than the Jeans length and
thus the growth rates obtained in our simulations are lower limits on
the physical growth rates. In Fig.~\ref{fig:structure} we visualize
various quantities of our $128^3$ resolution run to give an impression
of the complex structure within the central core.

\section{Conclusions}

Taken all together, our results strongly indicate that dynamically
important magnetic fields are generated during the formation of the
first stars. This has important consequences for our understanding of
how the first stars form and how they influence subsequent cosmic
evolution. We know from modeling galactic star-forming clouds that the
presence of magnetic fields can reduce the level of fragmentation, and
by doing so strongly influences the stellar mass spectrum \cite{HT08}.
Furthermore, we also know that dynamos in accretion disks produce jets
and outflows \cite{Rekowski03}, which remove a signiﬁcant fraction of
the mass and angular momentum. Again, this influences the stellar mass
spectrum. There are first attempts to study this process in the context
of first star formation \cite{Machida+06}, but more sophisticated
initial conditions and more appropriate magnetic field geometries need
to be considered. Once the first stars have formed, they are likely to
produce a copious amount of ionizing photons, which drive huge HII
regions, bubbles of ionized gas, expanding into the low-density gas
between the halos. These dynamics could be substantially different if
magnetic outflows drive a cavity-wave into the surrounding gas. The
magnetic field may further affect fluid instabilities near the
ionization front.

The mechanisms discussed  here are likely to work  not only during the
formation of the first stars,  but  in all  types of  gravitationally
bound,  turbulent  objects. Highly  magnetized  gas  is thus  expected
already in the first galaxies,  which subsequently needs to become more
coherent to form the typical field structures that are observed in the
present-day Universe.


\section*{Acknowledgments}

R.B. acknowledges fundings from the German Science Foundation (DFG)
through the grant BA 3706. S.S.~thanks the DFG for financial support
via the priority program 1177 ``Witnesses of Cosmic History: Formation
and Evolution of Black Holes, Galaxies and their Environment'' (grant
KL 1358/10). C.F.~has received funding from the European Research
Council under the European Community's Seventh Framework Programme
(FP7/2007-2013 Grant Agreement no.~247060) and from a Discovery
Projects Fellowship (DP110102191) of the Australian Research Council.
C.F.,~R.B.,~and R.S.K.~acknowledge subsidies from the
Baden-W\"urttemberg-Stiftung (grant P-LS-SPII/18) and from the German
Bundesministerium f\"ur Bildung und Forschung via the ASTRONET project
STAR FORMAT (grant 05A09VHA). D.S.~thanks for funding from the
European Community's Seventh Framework Programme (FP7/2007-2013) under
grant agreement No~229517. Supercomputing time at the
Forschungszentrum J\"ulich (projects hhd20 and hhd14) are gratefully
acknowledged. The software used in this work was in part developed by
the DOE-supported ASC / Alliance Center for Astrophysical
Thermonuclear Flashes at the University of Chicago.
Figure~\ref{fig:structure} was produced with the open-source
visualization software \textsc{visit}.


\end{document}